\documentclass[useAMS,usenatbib]{mn2e}
\usepackage{natbib}
\usepackage{color,graphicx} 
\usepackage{amsmath}        
\usepackage{amsfonts}       
\usepackage{amssymb}        
\usepackage{bm}         
\usepackage{framed}         
\usepackage{upgreek}
\usepackage{xspace}
\usepackage{hyperref}
\usepackage{array}
\author[Keane \& Walker]{E.F.~Keane$^{1}$ \& C.R.H.~Walker$^{2,1}$
  \\ $^{1}$ SKA Organisation, Jodrell Bank Observatory, Lower
  Withington, Macclesfield, Cheshire, SK11 9FT, UK. \\ $^{2}$ Jodrell
  Bank Center for Astrophysics, The University of Manchester, Alan
  Turing Building, Manchester, M13 9PL, UK. \\} \date{\today}
\title[FRB Injections]{Fast Radio Burst Injection Tests}

\begin{document}

\maketitle

\begin{abstract}

  Searches for fast radio bursts (FRBs) are underway at a growing
  number of radio telescopes worldwide. The sample size is now
  sufficient to enable many investigations into the population
  properties. As such, understanding the true sensitivity thresholds,
  effective observing time expended, survey completeness and parameter
  space coverage has become vital for calibrating the observed
  distributions. Recently the Molonglo FRB search team reported on
  their, as yet unique, efforts to inject synthetic FRB signals into
  their telescope data streams. Their results show 10 percent of
  injections being missed, even at very high signal-to-noise (S/N)
  ratios. Their pipeline employs components considered standard across
  several telescopes so that the result is potentially alarming. In
  this paper we present a further look at these missed injections. It
  is shown that all of the missed injections can be explained by
  combinations of the noise statistics, mis-labelling, overly harsh data
  analysis cuts, incorrect S/N calculations and radio frequency
  interference. There is no need to be alarmed.

\end{abstract}


\section{Introduction}
The science of fast radio bursts (FRBs) is growing at an extraordinary
pace. As the discovery rate increases~\citep{pbj+16} progress is being
made on several fronts, in particular localising sources to study
progenitors~\citep{bdp+19,ravi19} and investigating the possibilities
for using FRBs as cosmological
probes~\citep{keane18,caleb19}. Completeness of FRB searches, in some
search parameter sub-space, is often either implicitly assumed, or is
assumed to be calibrate-able in some way~\citep{liam19}. Often
building upon such assumptions are calculations of population
properties, e.g. their sky rates and brightness
distributions~\citep{SUPERB2,smb+18,jem+19}. However, end-to-end
testing of FRB search pipelines is typically not
performed. Additionally the local environments, and in particular the
radio frequency interference (RFI) characteristics, can be critical to
account for but as this is very difficult to do it is often not
attempted. Furthermore it is common to make assumptions about the
dispersion measure (DM) foreground subtraction accuracy and precision,
i.e. that it is perfect.

After initial work to test FRB search algorithms~\citep{kp15}, efforts
have expanded considerably of late (L. Connor, priv. comm.). Testing
the efficiency of FRB search pipelines in general has recently become
a topic of widening effort in the field, reflecting the welcome wider
trend to make research results generally more reproducible. Recently
\citet{Wael2019} presented discoveries of five FRBs made with the
UTMOST telescope in Australia. In this work they also presented the
first efforts at injecting synthetic FRB signals directly into the
telescope data streams. They examined properties of these injected
signals, as recovered by their search pipeline, to determine a metric
for the effectiveness of their FRB searches. They report that
approximately 10 percent of $\sim 2000$ injected FRBs are missed,
including many examples with signal-to-noise ratio (S/N) above
40. This result is potentially very concerning. Below the explanations
for this are investigated and identified.

\section{UTMOST FRB injections}
The UTMOST project is the name given to the upgrade of the Molonglo
Observatory Synthesis Telescope in New South Wales,
Australia~\citep{bjf+17}. The project currently focuses on two science
goals: pulsar timing and FRB searches. Originally a third focus, on
radio imaging, was pursued but this has received less attention since
the decision, in 2017, to park the telescope at certain fixed zenith
angles, and to observe in a drift scan mode. This \textit{modus
  operandi} significantly decreases the required maintenance of the
telescope and, with the sophisticated autonomous observing system that
is in place, it has little impact on pulsar timing and FRB searches.

As an FRB search system UTMOST has made 10
discoveries~\citep{UTMOST1,UTMOST2}, and now routinely dumps raw
Nyquist-rate voltages when FRBs are detected in real time in
low-resolution power (i.e. square-law detected) data. In addition to
the recent discoveries reported by \citet{Wael2019}, they present a
new and unique feature to their search system: the injection of
synthetic FRB test signals directly into the square-law detected
telescope data stream. When an FRB, real or synthetic, is discovered a
snippet of the telescope data is recorded to disk. The real time
search system used is built around one of the most commonly used FRB
search pipelines: \textsc{heimdall}\footnote{See for example
  \citealt{bbf10} and
  \texttt{https://sourceforge.net/projects/heimdall-astro/}}. For the
$\sim 2000$ synthetic FRB signal injections, \citet{Wael2019} reported
that approximately 10 percent are missed with no visible trend in the
missed sub-sample in either DM, pulse duration or S/N. The expected
ideal recovery would be considerably better than this and with
well-known expected trends in each of the aforementioned
parameters. Possible explanations for this might include: (i) issues
with their \textsc{heimdall}-based pipeline, either with
\textsc{heimdall} itself or the other parts of the processing chain;
(ii) the RFI environment; (iii) the stability of the telescope system;
and (iv) the classification system; these are investigated below in
\S~\ref{sec:why_missed}.

\section{Reasons for missed injections?}\label{sec:why_missed}
As well as discovering a large fraction of all known FRBs,
\textsc{heimdall} has been shown, at least by the limited testing it
has undergone, to be an accurate algorithm \citep{kp15}. The S/N it
reports for a given input signal of known shape, width and DM is as
expected for the best matches to these features, when the noise is
reasonably well-behaved, i.e. Gaussian and with no strong residual
RFI. This criterion is not met by some other commonly used algorithms,
which has resulted in detectable FRBs remaining undetected for some
time \citep{Crawford2016,Zhang2019,Keane2019}; there are likely more
such FRBs in public domain datasets. 

The top panel of Figure 1 shows the distribution of UTMOST FRB test
events as a function of their injected S/N (priv. comm. V. Gupta). The
criteria for an event to be considered detected is identical to that
used in \citet{Wael2019} and so the only difference from the left-most
panel of their Figure 4 is the histogram binning used; here bin widths
are 1-sigma in injected S/N. The over-arching picture one gets is that
a large number of injected FRBs, with high S/N, are missed by the
system. However, upon further investigation, it turns out that a large
number of these are ``false injections'' (priv. comm. V. Gupta). These
are FRB injections that were scheduled to occur, but were in fact not
searched for as observing did not occur as planned at those scheduled
times (for a variety of reasons). As such these have not been `missed'
as no pipeline ever searched for them. The bottom panel of Figure 1
shows the injected sample (i.e. showing only those injections that
were searched for, in green) and the subset of those that were missed
using the \citet{Wael2019} criteria (in orange). Furthermore the
missed distribution with one of the selection criteria removed is also
shown (in yellow). The criterion in question was one that rejected
detections with best estimated DM values that are offset from the
injected values by a factor $1/4$ of the estimated value. Such a cut
can remove long-duration low-DM events~\citep{cm03}; this is the case
for the very brightest injected event; it is detected by
\textsc{heimdall} at high S/N in an RFI-free data snippet but is
filtered out using this rule\footnote{We note that some of the events
  excluded due to the DM-related cut are in fact truly missed, but due
  to the presence of RFI.}. \textit{Ad hoc} filtering is common at
many radio telescopes, usually originating as a means to remove local
RFI. It is plausible that already-searched archival data, from a
number of telescopes, has had \textit{bona fide} astrophysical signals
missed due to such procedures.

At the detection threshold, by definition, one misses exactly half of
the injected pulses due to the noise distribution. Below the threshold
one misses more than half, and above the threshold one misses less
than half, in a manner described by the noise statistics, search
pipeline efficiencies and the RFI environment. As the UTMOST system
injects Gaussian pulses\footnote{See
  \texttt{https://github.com/vg2691994/Furby}}, but \textsc{heimdall}
searches for top-hat pulses, the maximum recoverable S/N is thus a
factor of $(\pi/(8\ln 2))^{1/4} \approx 0.87$ of the injected S/N
\citep{mc03}. The maximum recoverable S/N is also shown on the upper
abscissa in Figure 1. The 9-sigma threshold used is thus
\textit{effectively} a $10.4$-sigma threshold in \textit{injected} S/N
owing to the filter shape mis-match. Taking this into consideration
the theoretical expectation for the missed fraction is over-plotted
(in blue) in the lower panel of Figure 1. The number of injected
events per bin is too low to fully establish the noise
distribution\footnote{One would need several hundred FRBs injected in
  each 1-sigma histogram bin to establish the expected average S/N
  with a distribution whose rms equaled 1~\citep{kp15}.}, but with
this caveat in mind it seems that the first few bins agree with the
theoretical expectation. The expectation is also that no pulses should
be missed for injected S/N~$\gtrsim 13$. Missed detections for injected
S/N~$\lesssim 12$ are credibly explained as a combination of the
mis-match in pulse shape used in the injection and search, the
mis-match between true and trial values for DM and duration, as well as
the noise fluctuations in the data. With these considerations (and the
previous mis-labelling) only $\sim 1$\% of injections remain missed
without credible explanation, before one has examines the data or the
search pipeline specifics.

\begin{figure}
  \begin{center}
    \includegraphics[trim = 18mm 10mm 0mm 10mm, clip, scale=0.35,angle=0]{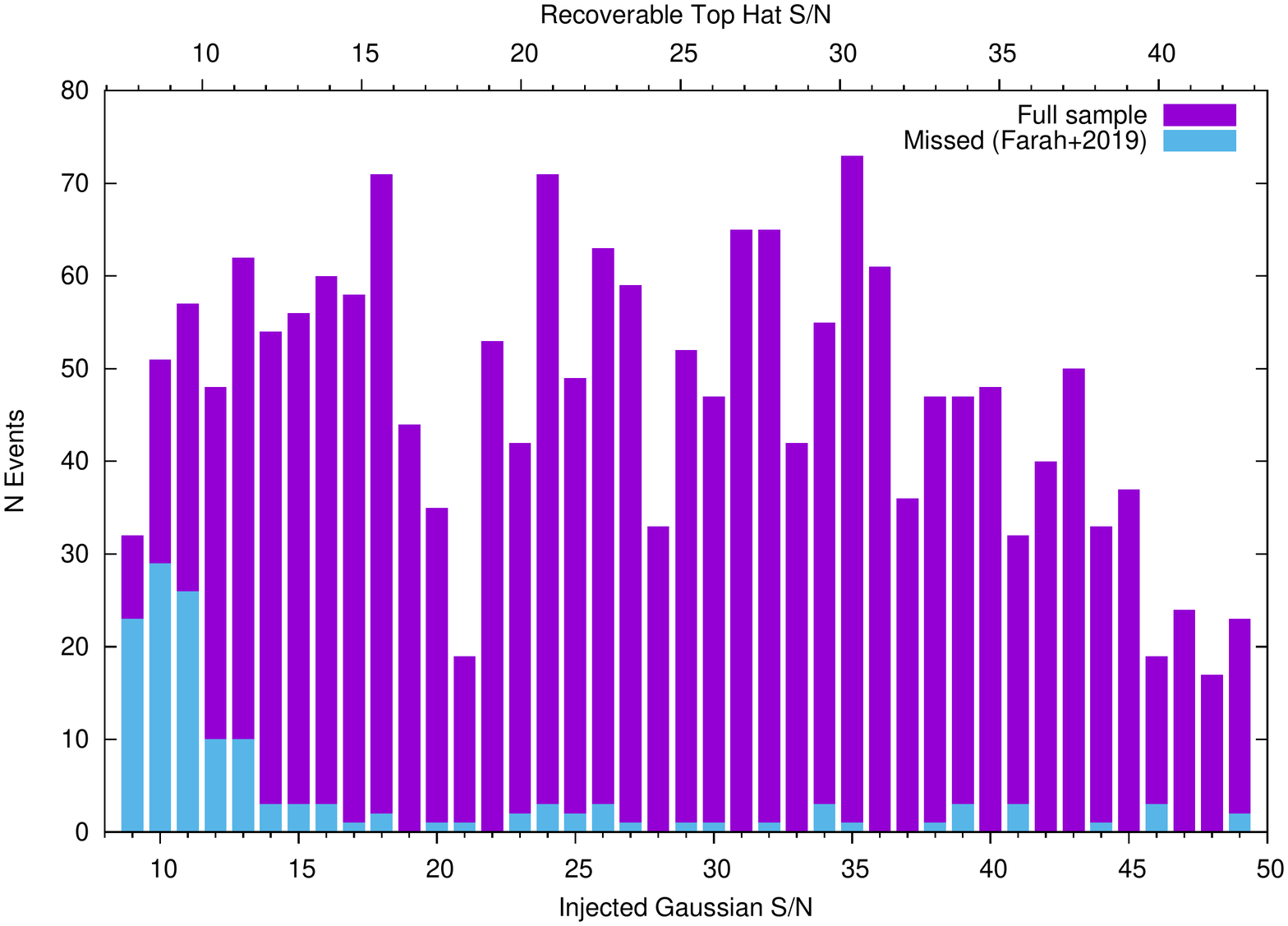}
    \includegraphics[trim = 18mm 10mm 0mm 10mm, clip, scale=0.35,angle=0]{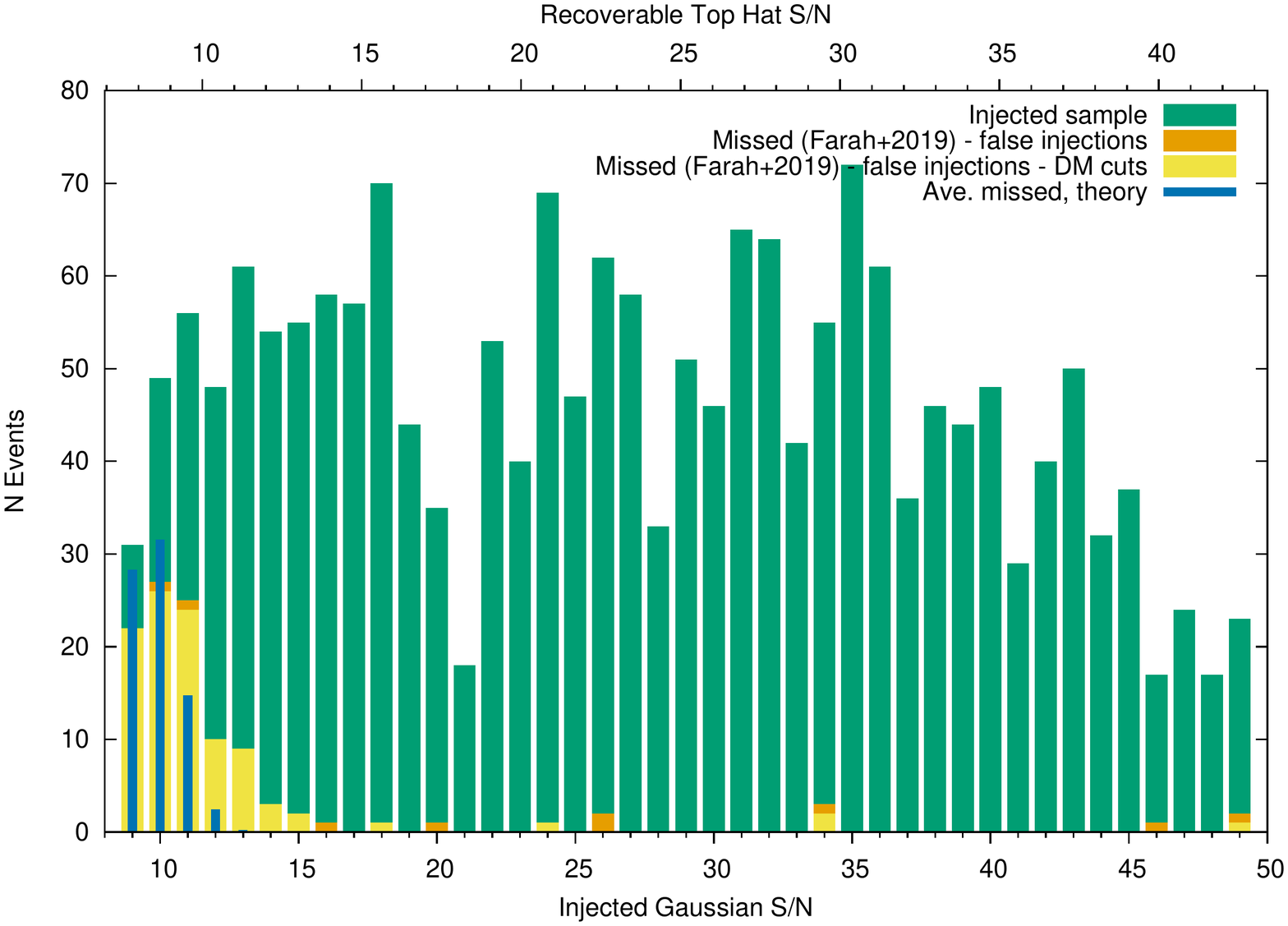}
    \caption{The top panel is a reproduction of Fig 4(a) from
      \citet{Wael2019}, re-binned to 1-sigma injected S/N bin
      widths. The bottom panel shows the sample which were actually
      injected (green), injected and searched (orange), and injected
      and searched with one selection cut, as discussed in the main
      text, removed (yellow). The theoretical expectation for the
      numbers missed, on average, is overplotted (blue). The latter
      highlights that missed events with injected S/N~$\gtrsim 13$
      require further scrutiny of the noise and RFI environments, and
      of particular corners of the search configuration space; doing
      this all missed injections are explainable.\label{fig:1}}
  \end{center}
\end{figure}

\section{The brightest missed injections}
Most of the missed 10 percent are explained as above, but the
remaining sample of missed injections, with the highest injected S/N
values, are however of the most concern; there are 10 missed FRBs with
injected S/N of $14$ or greater. This may be due to issues with the
search pipeline or the RFI/noise environment at the time of the
injections. The former would be correctable, whereas the latter may
not be. Fortunately a 2.95-second snippet of data, containing the true
sky noise at the time of each event, has been retained
(priv. comm. V. Gupta) and with this one can assess the noise and RFI
environment. Examining the missed events with injected S/N~$\geq 14$ by
eye shows that all but two cases can be explained by the presence of
strong unfiltered RFI in the $840-845$~MHz range. This information can
be used to iteratively `train' the pipeline, so as to improve it in
future to be even more effective. Such procedures are vital as the RFI
environments, even in the most remote observatory locations on Earth,
continue to get worse with time; RFI due to satellite communications
are particularly difficult to evade.

Of the two remaining signals: (i) the first missed event has injected
S/N~$=15$ and a pulse width of $\sim 2.5$ time samples implying a
maximum recoverable top-hat S/N of $\sim(15)(0.87)(2/2.5)^{0.5}\approx
11.7$ if the trial DM and trial pulse duration were perfectly matched
(which, unsurprisingly, they were not). From visual inspection no RFI
is evident and the pulse can \textit{just} be discerned, by eye, but
is credibly in a `trough' of the noise distribution; (ii) the second
has injected S/N~$=25$ and is quite obvious to the eye in an apparently
RFI-free snippet of data. The reason for its non-detection is thus
unclear at first, and this makes it the only concerning non-detection
in the sample. However, the data snippet containing this event has
some distinguishing properties which hint as to why it was missed. It
is the only incomplete data file in the entire sample of events
injected into the telescope data stream. The duration of the data
snippet is $<2.2$~s, in contradiction of its own data header. Its DM
value is $3025.7\;\mathrm{pc}\,\mathrm{cm}^{-3}$ and the low-frequency
end of the pulse is seen to be `chopped off' due to the very large
dispersion sweep across the band, and the reduced length of the
file. In combination these facts identify this event as unique and
potentially point at a reason why this injected event was missed.

\section{Conclusions \& Discussion}
In summary, it does not appear that $\sim10\%$ of the $\sim 2000$
simulated FRB signals recently injected into the UTMOST data stream
were missed. Over and above those that are consistent with noise
fluctuations, mis-labelling, overly harsh data cuts and the presence
of RFI, only 1 event was seen to be potentially suspicious. After
communicating the above to the UTMOST team, further investigations on
their part determined that \textsc{heimdall} had not actually been
employed to search for this $3025.7\;\mathrm{pc}\,\mathrm{cm}^{-3}$
event (V. Gupta, priv. comm.). The pipeline, designed to read the data
stream in `gulps' of time, was written such that when the last such
gulp was incomplete it discarded the data without searching it, and
without issuing an error/warning/report. This exact scenario happened
for the final remaining unexplained injected FRB.
Comfortingly then, the detection, or not, of all injected FRB signals
in the UTMOST data are explained. The analysis put forward here also
highlights the difficult nature, yet importance of, careful
consideration of the RFI environment and in the setting of search
pipeline parameters.

\subsection*{Acknowledgements}
EFK would like to thank and commend Vivek Gupta, Wael Farah and Chris
Flynn for sharing their data, injection search results, helpful and
instructive explanations and discussions. 
EFK would also like to thank Manisha Caleb for helpful discussions.

\bibliographystyle{mnras}

\end{document}